\documentclass[conference]{IEEEtran}
\IEEEoverridecommandlockouts
\usepackage{cite}
\usepackage{amsmath,amssymb,amsfonts}
\usepackage{graphicx}
\usepackage{textcomp}
\usepackage{xcolor}
\def\BibTeX{{\rm B\kern-.05em{\sc i\kern-.025em b}\kern-.08em
    T\kern-.1667em\lower.7ex\hbox{E}\kern-.125emX}}
\newcommand{\est}[1]{{\mathcal E}\!\left\{#1\right\}}

\newcommand{\rhant}{
\mbox{$ \; \circ\hspace*{-.5pt}\rule[2pt]{10pt}{.5pt}\!\bullet\; $}}
\newcommand{\lhant}{
  \mbox{$\; \bullet\!\rule[2pt]{10pt}{.5pt}\hspace*{-.5pt}\circ \; $}}

\newcommand{\CNormal}[2]{\mathcal{CN}(#1,#2)}

\newcommand{\Complex}[0]{\mathbb C}

\newcommand{\Vektor}[1]{{\mathbf{#1}}}
\newcommand{\uVektor}[1]{{\underline{#1}}}
\newcommand{\Matrix}[1]{{\mathbf{#1}}}
\newcommand{\PVektor}[1]{{\boldsymbol{#1}}}
\newcommand{\PMatrix}[1]{{\boldsymbol{#1}}}
\newcommand{\Herm}[0]{^{\mathrm{H}}}
\newcommand{\PH}[0]{^{\mathrm{P}}\!}
\newcommand{\Trans}[0]{^{\mathrm{T}}}

\newcommand{\diag}[1]{\mbox{diag}\!\left\{#1\right\}}

\newcommand{\ee}{\mathrm{e}}
\newcommand{\jj}{\mathrm{j}}
\newcommand{\mrt}{\mathrm{t}}
\newcommand{\ejo}{\ee^{\jj\Omega}}

\newif\ifShowCorrections
\ShowCorrectionsfalse
\ifShowCorrections
  \usepackage[normalem]{ulem}
  \definecolor{forgreen}{rgb}{0,0.6,0}
  \definecolor{orange}{RGB}{255,140,0}
  \definecolor{skyblue}{RGB}{100, 150, 235}
  
  \newcommand{\ikpc}[1]{{\color{forgreen}[#1]}}
  \newcommand{\sw}[2]{{\color{red}\sout{#1}}{\color{blue}#2}}
  \newcommand{\swc}[1]{{\color{orange}[#1]}}
\else
  
  \newcommand{\ikpc}[1]{}
  \newcommand{\sw}[2]{#2}
  \newcommand{\swc}[1]{}
\fi

\begin{document}

%
%
\title{Computational and Numerical Properties of a Broadband
  Subspace-Based Likelihood Ratio Test \thanks{Cornelius Pahalson
    acknowledges support by the Tertiary Education Trust Fund,
    Nigeria.}  }

\author{\IEEEauthorblockN{Cornelius A.D.~Pahalson,
    Louise H.~Crockett, and Stephan Weiss}
  \IEEEauthorblockA{\textit{Department of Electronic \& Electrical Engineering, University of Strathclyde,
    Glasgow, Scotland} \\
\{cornelius.pahalson,louise.crockett,stephan.weiss\}@strath.ac.uk}
}

\maketitle

%
%
\begin{abstract}
This paper investigates the performance of a likelihood ratio test in combination with a polynomial subspace projection approach to detect weak transient signals in broadband array data. Based on previous empirical evidence that a likelihood ratio test is advantageously applied in a lower-dimensional subspace, we present analysis that highlights how the polynomial subspace projection whitens a crucial part of the signals, enabling a detector to operate with a shortened temporal window. This reduction in temporal correlation, together with a spatial compaction of the data, also leads to both computational and numerical advantages over a likelihood ratio test that is directly applied to the array data. The results of our analysis are illustrated by examples and simulations.
\end{abstract}


%
%
\section{Introduction
  \label{sec:intro}}

A number of applications require the detection of weak transient
broadband signals in the presence of more dominant signals or
interference. This includes, for example, the task of detecting an
emerging primary user in a cognitive radio environment~\cite{quan08a,
  axell13a, alali23a, shukla07a, renard16a, perez22a, pahalson24a}, of
condition monitoring and and testing for electromagnetic
compatibility~\cite{tong16a}, or for registering seismic
events~\cite{ferber85a}. In a defence context, it is often desirable
if not vital to detect a weak transient source in
underwater/sonar~\cite{yin23a} or radio frequency domain
scenarios~\cite{weiss21b}.  Similarly, there may be a need to detect
the presence of a new speaker in an audio environment against several
stronger, overlapping speakers~\cite{neo22c} or against general
background noise~\cite{neo22b}.

The detection of transient signals can rely on energy-based criteria
and utilise short-time Fourier transform-type or wavelet-based
approaches to identify the correlation structure that transients may
be expected to possess~\cite{friedlander89a, friedlander92a,
  porat92a}.  Data-dependent transforms, for example the
Karhunen-Loeve transform~\cite{haykin91a}, reached via an eigenvalue
decomposition (EVD) of the data covariance matrix, can attain
optimality in terms energy compaction into a lower-dimensional
subspace.  Related subspace partitioning approaches have been used in
e.g.~\cite{scharf94a ,strobach94a, lundstrom98a, wang00b,
  wang01a}. More recently, machine learning methods have also been
attempted~\cite{yin23a, li23b}, but require a sufficient amount of data
in order to be trained off-line.

\begin{figure}[b]
  \includegraphics[width=\columnwidth]{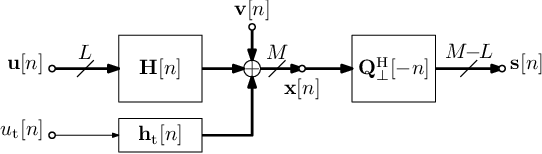}
  \caption{Signal model with measurement $\Vektor{x}[n]\in\Complex^M$,
    containing stationary sources $\Vektor{u}[n] \in\Complex^L$ and a
    transient source $u_{\mathrm{t}}[n]$, and processing by
    $\Matrix{G}[n]$ to yield a subspace projection $\Vektor{y}[n] \in
    \Complex^{M-L}$.
     \label{fig:model}}
\end{figure}

In order to address the problem of broadband transient signal
detection within a reasonable computational error, in the past we have
suggested a broadband or polynomial subspace
approach~\cite{weiss21b}. It is based on the assumption that $L$
sources that are stationary for a sufficient amount of time
illuminate $M$ sensors, with $M>L$. The propagation environment is
broadband, such that the sources arrive convolutively mixed,
i.e. ~posses both temporal and spatial correlation. In the presence of
these signals, we would like to detect the presence of a transient
signal. This scenario is outlined in Fig.~\ref{fig:model}; all signal
sources are assumed to be Gaussian, and we are only given the
measurements $\Vektor{x}[n]$ but are blind to the source model, with
the convolutive mixing systems that generate those measurements.

The statistically optimum test, the likelihood ratio test (LRT), in
the case of Gaussian data is based on the covariance of the data with
and without the transient signal~\cite{renard16a, besson20a,
  lehmann22a}. For broadband signals, temporal averaging in the LRT
needs to take the temporal correlation of the signals into account ---
this leads to potentially large space-time covariance matrices that
require inversion~\cite{pahalson24a}. For this reason,
in~\cite{pahalson24a}, the LRT has been combined with a relatively
inexpensive subspace approach, where the LRT is applied to a
lower-dimensional subspace projection rather than to the original
data. This was motivated by the fact that the subspace method itself
--- known for the narrowband case in~\cite{scharf94a, strobach94a,
  lundstrom98a, wang00b, wang01a} --- was successfully deployed for
weak transient signal detection~\cite{weiss21b, neo22b, neo22c}.

For the weak transient signal detection
in~\cite{weiss21b,neo22b,neo22c}, it is assumed that over a past
period of time, the statistics of the stationary signals can be
estimated.  Going forward in time, a change point detection would aim
to find any change in the signal energy.  This change point detection
has been easier to apply to a lower-dimensional vector
$\Vektor{s}[n]$, which is a projection onto the noise-only subspace of
the covariance matrix based on the stationary sources, than on the
measurements~$\Vektor{x}[n]$~\cite{weiss21b}.
For this approach, the particular statistics of the transient source
were not required to be known. However, in order to assess how close
this approach was to an optimal detection method, we applied an LRT
and a generalised LRT to both the measurements and the projected data
in~\cite{pahalson24a}, with significant empirical benefits for the
later.

Therefore, the aim of this paper is to explore why the polynomial
subspace approach is so beneficial for a combination with the
LRT. Particularly two issues have been noted in~\cite{pahalson24a}
with respect to the temporal window $T$, i.e.~basing a decision not
just on a single snap-shot but a sequence of $T$ time instances:
\begin{enumerate}
\item for small $T$, the subspace-based approach outperforms an LRT
  applied directly to the measurements;
\item for larger $T$, applying the LRT to the data rather than the
  projection can result in severe numerical problems.
\end{enumerate}
The contribution of this paper is to provide the theoretical
foundation to understand both points.  

Below, the system model, as well as the broadband subspace
decomposition and projection are outlined in Sec.~\ref{sec:cov}.
Sec.~\ref{sec:lrt} reviews the LRT method. The main contribution of
this paper is Sec.~\ref{sec:analysis}, which analyses the LRT when
applied to both measurement and projected data, followed by examples
and simulations in Sec.~\ref{sec:sims}.

%
%
\section{Space-Time Covariance and Polynomial Subspace Approach 
   \label{sec:cov}}

\subsection{Signal Model and Space-Time Covariance}

To expand on the signal model of Sec.~\ref{sec:intro} in
Fig.~\ref{fig:model}, we assume that in the stationary case,
contributions by $L$ sources are received by $M>L$ sensors via a
convolutive mixing matrix $\Matrix{H}[n] \in \Complex^{M\times L}$.
The sources, gathered in a signal vector $\Vektor{u}[n]
\in\Complex^L$, are mutually independent, temporally uncorrelated,
with zero mean and unit variance signals $u_\ell[n]$,
$\ell=1,\dotsc,L$, such that $\Vektor{u}[n] =
[u_1[n],\dotsc,u_L[n]]\Trans \sim \CNormal{\uVektor{0}}{\Matrix{I}_L}$.
Any specific source power spectral densities are captured in the model
of Fig.~\ref{fig:model} via the convolutive mixing matrix
$\Matrix{H}[n]$, which also operates as an innovation
filter~\cite{papoulis91a}. The data received from the sources is
corrupted by additive complex Gaussian noise $\Vektor{v}[n] \sim
\CNormal{\uVektor{0}}{\sigma^2_v \Matrix{I}_M}$. A transient signal
potentially illuminates the array via a vector of filters
$\Vektor{h}_{\mrt}[n]\in\Complex^M$ to another uncorrelated signal
$u_{L+1}[n] \sim \CNormal{0}{\sigma^2_t}$, whose presence we would like
to detect.

For the space-time covariance, based on the expectation $\est{\cdot}$
and Hermitian transpose operator $\{\cdot\}\Herm$, we have
$\Matrix{R}[\tau] = \est{\Vektor{x}[n]\Vektor{x}\Herm[n]}$ of the
measurement vector $\Vektor{x}[n]$. We first consider the absence of
the transient signal, i.e.~$u_{L+1}[n]=0$. In this case, for the
cross-spectral density (CSD) $\PMatrix{R}(z) = \sum_{\tau}
\Matrix{R}[\tau]z^{-\tau}$, or short $\Matrix{R}[\tau] \rhant
\PMatrix{R}(z)$, we obtain
\begin{align}
  \PMatrix{R}(z) = \PMatrix{H}(z)\PMatrix{H}\PH(z) + \sigma^2_{v}
  \Matrix{I}_M \;.
  \label{eqn:model1}
\end{align}  
In \eqref{eqn:model1}, $\PMatrix{H}(z) \lhant \Matrix{H}[n]$
represents a matrix of transfer functions. The parahermitian operator
$\{\cdot\}\PH$ implies a Hermitian transposition and time reversal,
such that $\PMatrix{H}\PH(z)=
\{\PMatrix{H}(1/z^\ast)\}\Herm$~\cite{vaidyanathan93a}. Hence,
$\PMatrix{R}(z)$ is a parahermitian matrix that satisfies
$\PMatrix{R}\PH(z)=\PMatrix{R}(z)$.

If $u_{L+1}[n]\neq 0$, then this signal's contribution to the overall
CSD matrix is
\begin{align}
  \PMatrix{R}_{\mrt}(z) = \PVektor{h}_{\mrt}(z)\PVektor{h}_{\mrt}\PH(z) \; ,
  \label{eqn:model1b}
\end{align}
with $\PVektor{h}_{\mrt}(z) \lhant \PVektor{h}[n]$. The overall CSD of
the measurement vector $\Vektor{x}[n]$ thus is $\PMatrix{R}(z) +
\PMatrix{R}_{\mrt}(z)$, whereby $\PMatrix{R}_{\mrt}(z)$ is a rank one
contribution.

\subsection{Analytic Eigenvalue Decomposition}

Given the signal model in Fig.~\ref{fig:model}, the CSD matrix
$\PMatrix{R}(z)$ is analytic in $z$ and admits an analytic
EVD~\cite{weiss18a, weiss18b, barbarino23a, weiss24b}
\begin{align}
  \PMatrix{R}(z) & = \PMatrix{Q}(z) \PMatrix{\Lambda}(z)
  \PMatrix{Q}\PH(z) \; .
  \label{eqn:analytic_evd}
\end{align}  
In \eqref{eqn:analytic_evd}, $\PMatrix{\Lambda}(z) =
\diag{\lambda_1(z),\dotsc,\lambda_M(z)}$ is a diagonal parahermitian
matrix holding the analytic eigenvalues $\lambda_m(z)$, $m=1,\dotsc,M$
of $\PMatrix{R}(z)$. Their corresponding eigenvectors form the columns
of $\PMatrix{Q}(z)$, which is also analytic in $z$ and satisfies
paraunitarity, such that $\PMatrix{Q}\PH(z)\PMatrix{Q}(z) =
\PMatrix{Q}(z)\PMatrix{Q}\PH(z) = \Matrix{I}_M$,and therefore
$\PMatrix{Q}^{-1}(z) = \PMatrix{Q}(z)$ result~\cite{vaidyanathan93a}.
Analyticity of the factors in \eqref{eqn:analytic_evd} is important,
as this property permits to approximate $\PMatrix{Q}(z)$ arbitrarily
closely by polynomials of sufficient order by shift and truncation
operations~\cite{weiss23b, corr15a,
  corr15b}.

Based on \eqref{eqn:analytic_evd}, we can define subspace decomposition,
\begin{align}
  \PMatrix{\Lambda}(z) & = \left[ \begin{array}{cc}
      \PMatrix{\Lambda}_{H}(z) + \sigma^2_v\Matrix{I}_{L} & \\
      &  \sigma^2_v \Matrix{I}_{M-L} \end{array} \right] \; ,
     \label{eqn:ss_L} \\
  \PMatrix{Q}(z) & = [\PMatrix{Q}_{\parallel}(z), \;\;
    \PMatrix{Q}_{\perp}(z)] \; ,
     \label{eqn:ss_Q}    
\end{align}  
where $\PMatrix{\Lambda}_{H}(z)$ holds the $L$ analytic eigenvalues of
$\PMatrix{H}(z)\PMatrix{H}\PH(z): \Complex \rightarrow
\Complex^{L\times L}$, and $\PMatrix{Q}_{\parallel}(z)$ contains their
corresponding eigenvectors. The orthogonal complement
$\PMatrix{Q}_{\perp}(z)$, such that
$\PMatrix{Q}_{\perp}\PH(z)\PMatrix{Q}_{\parallel}(z) = \Matrix{0}$,
contains analytic eigenvectors that span the noise-only subspace of
$\PMatrix{R}(z)$. This noise-only subspace does not contain any
contributions by the $L$ stationary sources.

In practice, analytic EVD algorithms~\cite{tohidian13a, weiss19a,
  weiss21a, weiss20c, weiss23b, khattak24b, khattak24d} operate in the
DFT domain. In case the eigenvalues are spectrally majorised,
e.g.~because of estimation~\cite{delaosa18a, delaosa19a, khattak22c,
  bakhit24a}, polynomial EVD algorithms~\cite{mcwhirter07a, redif11a,
  redif11b, redif15a, corr14a, corr14d, wang15a, mcwhirter16a} can
yield similar results. Computationally efficient implementations of
such decompositions have been considered in
e.g.~\cite{coutts19a,khattak21a}; the order of $\PMatrix{Q}(z)$ ---
which will determine the order and therefore computational cost of
filters in applications --- can be reduced through limiting its order
through shifts and truncations~\cite{foster06a, ta07d, corr15a,
  corr15b, weiss23b}.

\subsection{Subspace Projection 
  \label{sec:syndrome}}

With the partitioning of the analytic SVD factors in \eqref{eqn:ss_L}
and \eqref{eqn:ss_Q}, $\PMatrix{Q}_{\perp}(z) \lhant \Matrix{Q}_{\perp}[n]$
can be used to project the measurement data $\Vektor{x}[n]$ into the
noise-only subspace,
\begin{align}
  \Vektor{s}[n] & = \sum_\nu \Matrix{Q}\Herm_{\perp}[-\nu]
     \Vektor{x}[n-\nu] \; ,
  \label{eqn:syndrome}
\end{align}
where $\Vektor{s}[n] \in \Complex^{M-L}$ is the projected data, as
shown in Fig.~\ref{fig:model}.  In the ideal case, this new signal
vector $\Vektor{s}[n]$ is now free of any contributions from the $L$
stationary sources. In contrast, with the emergence of the transient
signal $u_{L+1}[n]$, there are now $L+1$ signals in the environment,
and at least some part of $u_{L+1}[n]$ will project into
$\Vektor{s}[n]$, where its presence can be more easily detected than
in the measurements $\Vektor{x}[n]$.

Because part of the broadband signal $u_{L+1}[n]$ is projected into
the noise-only subspace vector $\Vektor{s}[n]$, the latter has also
been termed a syndrome vector in~\cite{weiss21b}. This terminology is
borrowed from coding theory and in particular from filter-bank based
source-channel coding methods~\cite{labeau05b}, where a broadband
subspace that is orthogonal to the code subspace is indicative of
impulse noise and other transmission errors. The operation in
\eqref{eqn:syndrome} represents a generalisation of narrowband
subspace detection approaches~\cite{scharf94a}; the extension to the
broadband case via \eqref{eqn:syndrome} has been utilised for voice
activity detection in the presence of stronger speakers~\cite{neo22c}
or noise~\cite{neo22b}.
Such polynomial subspace decompositions have also been applied, for
example, in the context of joint source-channel
coding~\cite{weiss06a}, angle of arrival
estimation~\cite{alrmah11a,weiss13a,hogg21a}, source
separation~\cite{redif17a} and localisation~\cite{neo23b},
beamforming~\cite{weiss15a, neo23a}, and channel
identification~\cite{weiss17a}.

%
%
\section{Likelihood Ratio Test
  \label{sec:lrt}}

We now follow the approach in~\cite{pahalson24a}, where a likelihood
ratio approach can be applied to either the measurement data
$\Vektor{x}[n]$ or to its subspace projection $\Vektor{s}[n]$. We
first briefly comment on the LRT formulation before we focus on the
two application cases.

\subsection{Likelihood Ratio Test}

For a general exploration of the likelihood ratio test, we utilise a
test variable $\Vektor{y}_n\in\Complex^K$, which can later be
constructed from measurement vectors $\Vektor{x}[n]$ or syndrome
vectors $\Vektor{s}[n]$, including a concatenation of temporal
snapshots.  The dimension $K$ will therefore depend on this choice.
We assume that $\Vektor{y}_n$ can consist of $\Vektor{y}_{0,n} \in
\Complex^K$, which is the stationary noise, and
$\Vektor{y}_{1,n}\in\Complex^K$, representing the transient
component. Both signals are assumed to be zero mean complex Gaussian
with $\Vektor{y}_{0,n} \sim \CNormal{\uVektor{0}}{\Matrix{R}_0}$ and
$\Vektor{y}_{1,n} \sim \CNormal{\uVektor{0}}{\Matrix{R}_1}$. The aim
is thus to distinguish between the two hypotheses
\begin{align}
  H_0: & \quad \Vektor{y}_n = \Vektor{y}_{0,n} \nonumber \; , \\
  H_1: & \quad \Vektor{y}_n = \Vektor{y}_{0,n} + \Vektor{y}_{1,n} \nonumber \; .
\end{align}
In our context, $\Vektor{y}_{0,n}$ holds the contribution from the $L$
stationary sources and additive noise, while $\Vektor{y}_{1,n}$
comprises of components due to the transient signal.  The probability
density functions for $\Vektor{y}_n$, under the two hypotheses are
then given by
\begin{align}
  p(\Vektor{y}_n|H_0) & = (2\pi |\Matrix{R}_0|)^{-\frac12}
      \ee^{-\frac12 \Vektor{y}\Herm_n \Matrix{R}_0^{-1} \Vektor{y}_n} \; ,\\
  p(\Vektor{y}_n|H_1) & = (2\pi |\Matrix{R}_0+\Matrix{R}_1|)^{-\frac12}
       \ee^{-\frac12 \Vektor{y}\Herm_n (\Matrix{R}_0+\Matrix{R}_1)^{-1} \Vektor{y}_n} \; ,
\end{align}
where the determinant of a matrix $\Matrix{X}$ is denoted as
$|\Matrix{X}|$.

For the likelihood ratio $L(\Vektor{y}_n)$, we have
\begin{align}
  L(\Vektor{y}_n) & = \frac{p(\Vektor{y}_n|H_0)}{p(\Vektor{y}_n|H_1)}
   = \frac{|\Matrix{R}_0+\Matrix{R}_1|^{\frac12}}{|\Matrix{R}_0|^\frac12}
      \ee^{-\frac12 \Vektor{y}\Herm_n \Matrix{A} \Vektor{y}_n} \; ,
\end{align}
where for brevity
\begin{align}
  \Matrix{A} = \Matrix{R}^{-1}_0 - (\Matrix{R}_0 + \Matrix{R}_1)^{-1}  \;.
  \label{eqn:R_inversions}
\end{align}
With its EVD $\Matrix{A} = \Matrix{Q} \Matrix{\Lambda}
\Matrix{Q}\Herm$, the likelihood ratio can be further expressed
as
\begin{align}
  L(\Vektor{y}_n) & =  \frac{|\Matrix{R}_0+\Matrix{R}_1|^{\frac12}}
                          {|\Matrix{R}_0|^{\frac12}}
  \ee^{-\frac12 \| \Matrix{\Lambda}^{\frac12}\Matrix{Q}\Herm\Vektor{y}_n\|^2_2} \; .
\end{align}
In order to accept or reject the hypothesis, we now need to find a
threshold $c$ such that
\begin{align}
   L(\Vektor{y}_n) \overset{H_0}{\underset{H_1}{\lessgtr}} c  \; .
\end{align}
Rearranging leads to
\begin{align}
  \|\Matrix{\Lambda}^{\frac12}\Matrix{Q}\Herm\Vektor{y}_n\|
  \overset{H_0}{\underset{H_1}{\lessgtr}}
  2 \ln\left\{ \frac{|\Matrix{R}_0|^{\frac12}}
             {|\Matrix{R}_0+\Matrix{R}_1|^{\frac12}} c\right\} = c^\prime \; ,
  \label{eqn:test}
\end{align}
where $c^\prime$ is a modified threshold.  With the term
$\|\Matrix{\Lambda}^{\frac12}\Matrix{Q}\Herm\Vektor{y}_n\|$, we have
defined the test statistic for the likelihood ratio test.

\subsection{Likelihood Ratio Test on Measurements
   \label{sec:LRTx}}

Applying the LRT directly to the measurement data in principle
exploits all information that we can possibly draw from it. For the
purpose of temporal averaging the test outcome, for the LRT variable
$\Vektor{y}_n$ we utilise a concatenation of $T$ snapshots of
$\Vektor{x}[n]$,
\begin{align}
  \Vektor{y}\Herm_n = [\Vektor{x}\Herm_n, \Vektor{x}\Herm_{n-1},\dotsc,
    \Vektor{x}\Herm_{n-T+1}]  \; ,
  \label{eqn:concat_x}
\end{align}
such that $\Vektor{y}_n \in \Complex^{MT}$.

For the covariance matrices in \eqref{eqn:R_inversions}, we introduce
extra subscripts to denote their reference to the measurement vector
via $\Matrix{R}_{\Vektor{x},0}$ and $\Matrix{R}_{\Vektor{x},1}$. For
the covariance covering the hypothesis $H_0$, we have
\begin{align}
  \Matrix{R}_{\Vektor{x},0} & = \left[ \begin{array}{ccc}
      \Matrix{R}[0] & \hdots & \Matrix{R}[T-1] \\
      \vdots & \ddots & \vdots \\
      \Matrix{R}[1-T] & \hdots & \Matrix{R}[0]
    \end{array} \right] \; ,
  \label{eqn:Rz_to_R}
\end{align}
where \eqref{eqn:model1} defines $\Matrix{R}[\tau] \rhant
\PMatrix{R}(z)$.  Similarly, $\Matrix{R}_{\Vektor{x},1}$ can be
constructed from the lag components of $\Matrix{R}_\mrt[\tau]$ in
\eqref{eqn:model1b}.  Both covariance matrices
$\Matrix{R}_{\Vektor{x},i}$, $i=0,1$, are of dimension $(MT)\times
(MT)$.  Thus the inversions required for \eqref{eqn:R_inversions} can
present computational and numerical challenges, which we will address
separately in Sec.~\ref{sec:analysis}.

\subsection{Likelihood Ratio Test on Subspace Data
   \label{sec:LRTs}}

In order to apply the likelihood ratio test to $T$ successive
samples of the projected data vector $\Vektor{s}[n]$, we define
analogously to \eqref{eqn:concat_x} the variable
\begin{align}
  \Vektor{y}\Herm_n = [\Vektor{s}\Herm_n, \Vektor{s}\Herm_{n-1},\dotsc,
    \Vektor{s}\Herm_{n-T+1}]  \; ,
  \label{eqn:vec_y_s}
\end{align}
where now $\Vektor{y}_n \in\Complex^{(M-L)T}$.
For its space-time covariance $\Matrix{R}^\prime[\tau] =
\est{\Vektor{y}[n] \Vektor{y}\Herm[n-\tau]}$ and the corresponding CSD
matrix $\PMatrix{R}^\prime(z): \Complex \rightarrow
\Complex^{(M-L)\times(M-\sw{l}{L})}$, we can state 
\begin{align}
  \PMatrix{R}^\prime(z) & = \PMatrix{Q}\PH_{\perp}(z) \PMatrix{R}(z)
  \PMatrix{Q}_{\perp}(z) \; ,
  \label{eqn:Rs0z}
\end{align}
with respect to the CSD matrix of the measurement data,
$\PMatrix{R}(z)$, in Sec.~\ref{sec:LRTx}.
Analogous to \eqref{sec:LRTx}, we can now obtain
$\Matrix{R}_{\Vektor{s},0}$ from the matrix-valued coefficients of
$\Matrix{R}^\prime[n] \rhant \PMatrix{R}^\prime(z)$. 

For the transient component, we can define 
\begin{align}
  \PMatrix{R}^\prime_{\mrt}(z)  & = \PMatrix{Q}\PH_{\perp}(z) \PMatrix{R}_{\mrt}(z)
  \PMatrix{Q}_{\perp}(z) \; ,
     \label{eqn:Rprimet}
\end{align}
based on the rank-one term $\PMatrix{R}_{\mrt}(z)$ in
\eqref{eqn:model1b}. Analogously to \eqref{sec:LRTx}, we can obtain a
constant covariance matrix $\Matrix{R}_{\Vektor{s},0}$ from
$\Matrix{R}^\prime_{\mrt}[n] \rhant \PMatrix{R}^\prime_{\mrt}(z)$.
Overall, in the subspace-based case we now have covariance matrices
$\Matrix{R}_{\Vektor{s},i}$, $i=0,1$, of size $T(M-L) \times T(M-L)$.

\subsection{Application and Generalised LRT}

The application of the above LRT test to data typically assumes that
the covariance matrices under the two hypothesis are known ---
$\Matrix{R}_0$ for $H_0$, and the composite $(\Matrix{R}_0 +
\Matrix{R}_1)$ for $H_1$ --- independent of whether these are derived
from the measurements or the subspace-projected data. The covariance
matrices represent the exact ensemble statistics, i.e.~the source
model of Fig.~\ref{fig:model} is known a priori.  In practise, where
only finite data is available to estimate the statistics, an estimated
space-time covariance $\hat{\PMatrix{R}}(z)$ will be subject to
estimation errors that depend on both the sample size of the data, as
well as on the ground truth $\PMatrix{R}(z)$~\cite{delaosa18a,
  delaosa19a, khattak22c, bakhit24a}, thus resulting in estimates
$\hat{\Matrix{R}}_0$ and $(\hat{\Matrix{R}}_0 + \hat{\Matrix{R}}_1)$.
This reliance on potentially inaccurate estimates turns the LRT into a
generalised likelihood ratio test (GLRT).

Recall that the aim of this paper is to explore the optimality of the
test. For our application outlined in Sec.~\ref{sec:intro}, we have an
estimate $\hat{\Matrix{R}}_0$, but we are not able to measure
$\hat{\Matrix{R}}_1$, neither by itself or in combination with
$\hat{\Matrix{R}}_0$, ahead of performing a change point
detection. Knowing what is optimally achievable given either
$\Matrix{R}_i$ or $\hat{\Matrix{R}}_i$ provide a useful benchmark
for~\cite{weiss21b} and applications such as~\cite{neo22b, neo22c}.

%
%
\section{Analysis of Subspace-Based LRT
   \label{sec:analysis}}

In this section we explore properties of the covariance matrices
$\Matrix{R}_{\Vektor{x},i}$, $i=0,1$, for the measurements and of
$\Matrix{R}_{\Vektor{s},i}$, $i=0,1$, for the projected data, that
feed into the LRT discussed in Sec.~\ref{sec:lrt}.

\subsection{Consideration of Temporal Correlation}

We first focus on the hypothesis $H_0$, where only $L$ stationary
signals are present. In this case, if the mixing system
$\PMatrix{H}(z)$ consists of FIR filters of length $(J+1)$, i.e.~it is a
polynomial matrix of order $J$, then $\PMatrix{R}(z)$ is a Laurent
polynomial matrix of order $2J$. Generally, this matrix will be dense
in the sense that generally all its coefficients $\Matrix{R}[\tau]$
for $|\tau| \leq J$ will have non-zero elements. As a result, for the
consideration of a temporal window $T<J$ in the LRT,
$\Matrix{R}_{\Vektor{x},0}$ will be a dense matrix. Only for the case
$T>J$ will we start to see zero corner blocks to appear.

Inspecting the projected data under hypothesis $H_0$, we have
$\PMatrix{R}^\prime(z)$ given by \eqref{eqn:Rs0z}. With the subspace
partitioning in \eqref{eqn:ss_L} and \eqref{eqn:ss_Q}, this simplifies
in the case of an ideal EVD to
\begin{align}
  \PMatrix{R}^\prime(z) = \sigma^2_v \Matrix{I}_{M-L} \; ,
\end{align}  
since the projection we have orthogonality of the signal subspace,
i.e.~$\PMatrix{Q}\PH_{\!\perp}(z) \PMatrix{H}(z) = \Matrix{0}$. Thus,
for the LRT test we obtain
\begin{align}
  \Matrix{R}_{\Vektor{s},0} & = \sigma^2_v \Matrix{I}_{(M-L)T} \; .
  \label{eqn:Rs0_analysed}
\end{align}  
In contrast to $\Matrix{R}_{\Vektor{x},0}$ for the measurement data,
\eqref{eqn:Rs0_analysed} shows that under $H_0$, the test variable
$\Vektor{y}_n$ in \eqref{eqn:vec_y_s} is spatially and temporally
uncorrelated. Typically temporal correlation will degrade a test
variable~\cite{weiss21b} which is avoided for the projected data under
$H_0$.

Under hypothesis $H_1$, we have the covariance matrix $(\Matrix{R}_0 +
\Matrix{R}_1)$.  For both the measurement case and the projected data
case, $\Matrix{R}_{\Vektor{x},1}$ and $\Matrix{R}_{\Vektor{s},1}$ will
now be dense matrices, and the input data to the LRT, $\Vektor{y}_n$,
in both cases will be temporally and spatially correlated. Due to
passing through the filter bank $\PMatrix{Q}\PH_{\!\perp}(z)$, the
projected data will be correlated over an even longer data window
compared to the measurement data.

To build the matrix $\Matrix{A}$ in \eqref{eqn:R_inversions} for the
LRT, consider that both $\PMatrix{R}_{\mathrm{t}}(z)$ and
$\PMatrix{R}^{\prime}_{\mathrm{r}}(z)$ are rank one matrices as
evident from the outer products in \eqref{eqn:model1b} and
\eqref{eqn:Rprimet}, whereby for the latter we can define
$\PMatrix{R}^\prime_{\mrt}(z) = \PMatrix{Q}\PH_{\!\perp}(z)
\PVektor{h}_{\mrt}(z) \PVektor{h}\PH_{\mrt}(z) \PMatrix{Q}_{\perp}(z)
= \PVektor{h}^\prime_{\mrt}(z) \PVektor{h}^\prime{}\PH_{\mrt}(z)$ with
the $\PVektor{h}^\prime_{\mrt}(z)=\PMatrix{Q}\PH_{\!\perp}(z)
\PVektor{h}_{\mrt}(z): \Complex \rightarrow \Complex^{M-L}$ a vector
of functions. As a result of the space-time covariance having rank
one, at least theoretically the covariance matrices
$\Matrix{R}_{\Vektor{x},1}$ and $\Matrix{R}_{\Vektor{s},0}$ will have
at most rank $T$~\cite{neo23a} and can be factorised as
\begin{align}
  \Matrix{R}_{\Vektor{x},1} & = \Matrix{H}_{\mrt} \Matrix{H}\Herm_{\mrt} \\
  \Matrix{R}_{\Vektor{s},1} & = \Matrix{H}^\prime_{\mrt} \Matrix{H}^\prime{}\Herm_{\mrt}
\end{align}
where $\Matrix{H}_{\mrt} \in \Complex^{MT\times T}$ and
$\Matrix{H}^\prime{\mrt} \in \Complex^{(M-L)T\times T}$.

Using the Woodbury identity for the low-rank update $(\Matrix{R}_0 +
\Matrix{R}_1)^{-1}$, for \eqref{eqn:R_inversions} we obtain in
general
\begin{align}
  \Matrix{A} & = 
  \Matrix{R}_0^{-1}\Matrix{H}_{\mrt}(\Matrix{I}_T +
  \Matrix{H}\Herm_{\mrt}\Matrix{R}_0^{-1} \Matrix{H}_{\mrt})^{-1}
  \Matrix{H}\Herm_{\mrt}\Matrix{R}_0^{-1} \; .
  \label{eqn:Ax}
\end{align}
When \eqref{eqn:Ax} specifically for the measurement case, then the
dense nature of $\Matrix{R}_{\Vektor{x},0}$ does not allow for further
simplifications.  In the case of the projected data, here referred to
as $\Matrix{A}_{\Vektor{s}}$, we have
\begin{align}
  \Matrix{A}_{\Vektor{s}} & = \frac{1}{\sigma^2_v} \Matrix{H}^\prime_{\mrt}
  ( \sigma^2_v \Matrix{I}_T +
  \Matrix{H}^\prime{}\Herm_{\mrt}\Matrix{H}^\prime_{\mrt} )^{-1} 
  \Matrix{H}^\prime{}\Herm_{\mrt} \; .
  \label{eqn:As}
\end{align}
Although \eqref{eqn:Ax} and \eqref{eqn:As} do not represent how the
processor for the test statistic in \eqref{eqn:test} is computed since
only $\Matrix{R}_0$ and $(\Matrix{R}_0 + \Matrix{R}_1)$ are available,
\eqref{eqn:As} provides some insight into how the projected data case
simplifies the underlying procedure --- it provides an outer product
between a low rank matrix $\Matrix{H}^\prime_{\mrt}$ and its
regularised left pseudo-inverse, that is free of any temporal or
spatial correlations imposed by the stationary sources captured in
$\Matrix{R}_0$.

\subsection{Consideration of Covariance Matrix Conditioning
  \label{sec:cond}}

We now want to assess how the computation of the inverses for
$\Matrix{R}_0$ and $(\Matrix{R}_0 + \Matrix{R}_1)$ are affected by
whether we operate the LRT on the measurements or the projected data.
We assess this via the condition number of a matrix~\cite{golub96a},
which assesses the gain w.r.t.~any random perturbations ---
e.g.~through estimation errors~\cite{delaosa19a} --- in the inversion
process.

For the measurement case, the condition number of
$\Matrix{R}_{\Vektor{x},0}$ can be related to the eigenvalue
spread~\cite{haykin91a} of the signals, which assesses the ratio
between the minimum and maximum spectral value and here additionally
has a spatial component. For the maximum eigenvalue, we have
\begin{align}
  \max\{\lambda_{\Vektor{x},0}\} & = \max_{\Omega,m}
  \lambda\{\PMatrix{R}(\ejo)\} \\ & \approx \max_{\Omega,m}
     \lambda\{\PMatrix{H}(\ejo) \PMatrix{H}\PH(\ejo)\} 
      > \sigma^2_s \; ,
\end{align}
where $\lambda\{\PMatrix{R}(z)\}$ returns the analytic eigenvalues of
the parahermitian matrix $\PMatrix{R}(z)$~\cite{weiss18a}, and
$\sigma^2_s$ is the maximum power of a stationary source in a measurement
$x_m[n]$, assuming for the SNR $\sigma^2_s/ \sigma^2_v \gg 1$.  For
the smallest eigenvalue, the noise floor in the noise-only subspace
will be given by $\sigma^2_v$, such that we obtain
\begin{align}
    \gamma_{\Vektor{x},0} > \sigma^2_s/ \sigma^2_v
\end{align}  
for the condition number under $H_0$. Under $H_1$, the maximum
eigenvalue is still due to a strong stationary component, while the
minimum possible eigenvalue remains given by the noise floor. Hence,
we have approximately $\gamma_{\Vektor{x},1} = \gamma_{\Vektor{x},0}$.

For the case of projected data, ideally the stationary signals are no
longer present. In this case, the maximum eigenvalue is given by the
\begin{align}
  \max\{\lambda_{\Vektor{s},0}\} & = \max_{\Omega,m}
  \lambda\{\PMatrix{R}^\prime(\ejo)\} \\
  & = \max_{\Omega,m}
  \lambda\{\PMatrix{H}_{\mrt}(\ejo) \PMatrix{H}\PH_{\mrt}(\ejo)
      + \sigma^2_v \Matrix{I}_{T(M-L)} \} \nonumber \\
     & >  \sigma^2_{\mrt} + \sigma^2_v\; ,
\end{align}
where $\sigma^2_{\mrt}$ represents the maximum power of the transient
signal in any of the measurement signals $x_m[n]$. Note that this
power is not increased when passing through $\PMatrix{Q}\PH(z)$, since
this matrix completes to a paraunitary and therefore energy-preserving
system.  Since the smallest eigenvalue is still limited by the noise
floor, we therefore have $\gamma_{\Vektor{s},0} > 1$ and
\begin{align}
    \gamma_{\Vektor{s},1} > \frac{\sigma^2_t + \sigma^2_v}{\sigma^2_v}
\end{align}  
for the condition number under hypotheses $H_0$ and $H_1$. Note that
in the case that the transient signal is significantly weaker than the
stationary signals, the covariance matrices for the projected data
have a much lower bound for the condition number and generally will be
much better conditioned than their equivalent quantities based on the
measurement data.

%
%
\section{Simulations and Results
   \label{sec:sims}}

\subsection{System Setup and Performance Metrics
  \label{sec:metric}}

To demonstrate the LRT and its analysis, with reference to
Fig.~\ref{fig:model} we investigate a scenario where we have $M=10$
sensors picking up signals from $L=7$ independent stationary sources
via a mixing system $\PMatrix{H}(z)$ of two different orders $J=\{10;
20\}$. The SNR of these signal is 20~dB w.r.t.~additive complex-valued
uncorrelated Gaussian noise of variance $\sigma^2_v$.  A transient
source is undergoing a source model of the same order as
$\PMatrix{R}(z)$ and at the sensors possess as powers that are $\{10;
20\}$~dB below the stationary signals, i.e.~in the second setting the
transient signal sits in the noise floor. The mixing system is
generated via source power spectral density models and a paraunitary
mixing system, such that the ground truth space time covariance matrix
$\PMatrix{R}(z)$ is known for the LRT. It is also estimated from
$10^5$ snapshots of data using the best linea unbiased estimator
in~\cite{delaosa19a} for GLRT results.

To compare the different methods, a good metric for the separation of
distributions is the
receiver operating characteristic (ROC)~\cite{hanley82a}. Here instead we
work with scalar metric 
\begin{align}
  \delta & = \frac{|\mu_1 - \mu_0|}{(\sigma_0+\sigma_1)/2} \;.
  \label{eqn:separability}
\end{align}
which define the separation distance between the distributions under
hypotheses $H_0$ and $H_1$. This metric $\delta$ assesses the ratio
between the distributions' means $\mu_i$ , normalised by the mean of
their standard deviations $\sigma_i$, for the two hypotheses $H_i$,
$i=0,1$.

\subsection{Simulations
  \label{sec:basic_sim}}

\begin{figure}
  \hspace*{.15cm}  \includegraphics[width=0.975\columnwidth]{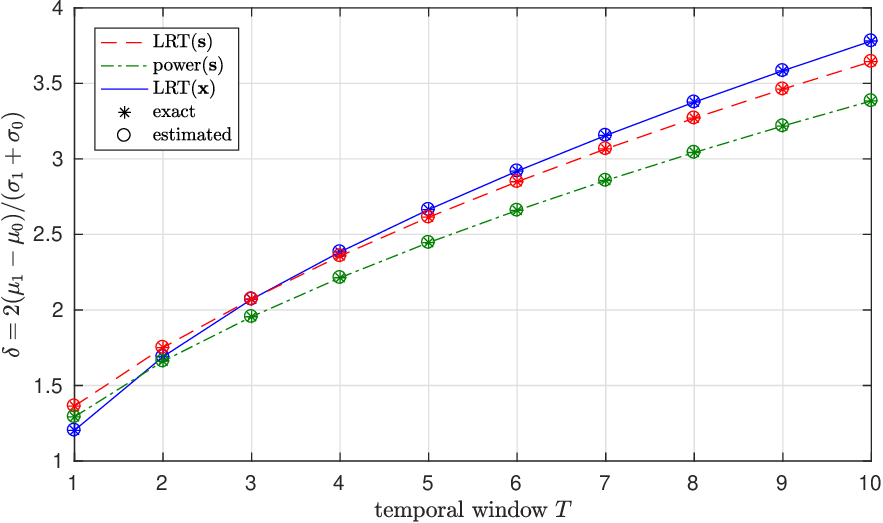}

  \vspace*{-.3cm}
  
  \caption{Separability of distributions for setting with $J=10$ and
    the transient source 10~dB below the stationary sources.
    \label{fig:fig3}}
\end{figure}

For the setting with $J=10$ and the transient signal sitting 10dB
below the stationary signal in power, the results for the separability
$\delta$ as defined in~\eqref{eqn:separability} is shown in
Fig.~\ref{fig:fig3} as a function of the temporal window $T$, $1 \leq
T \leq 10$. For small values of $T$, decorrelating property of
$\PMatrix{Q}\PH_{\perp}(z)$ give an advantage to the LRT operating on
the projected data. Even just assessing the power of the projected
data averaged over $T$ snapshot without taking temporal correlation
into account, as exploited in~\cite{weiss21b,neo22b, neo22c} and
marked as power($\Vektor{s}$) in Fig.~\ref{fig:fig3}, provides and
advantage over the LRT directly applied to the measurement data. Only
as $T$ is increased will the LRT of the measurement data outperform
the other approaches, due to it additionally exploiting any
information on the transient source that resides within the signal
plus noise subspace.

  
\begin{figure}

  \hspace*{.15cm} \includegraphics[width=0.975\columnwidth]{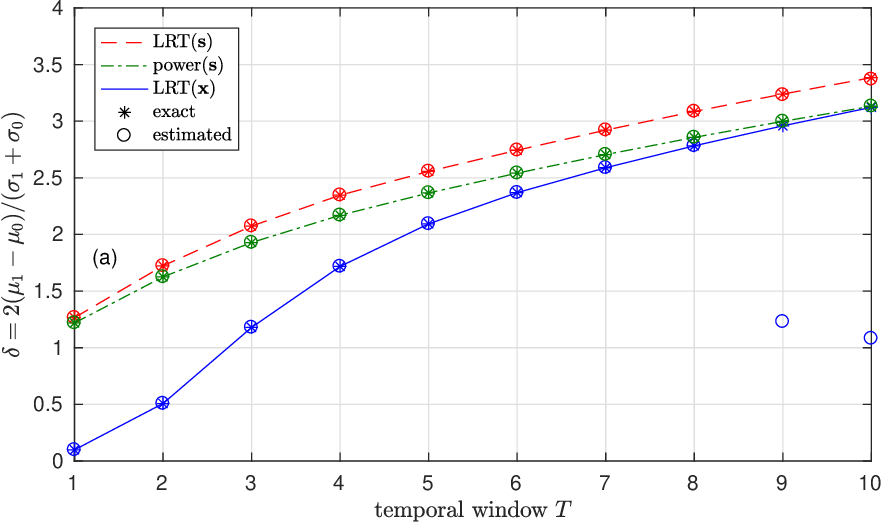}

  \vspace*{.2cm}

  \includegraphics[width=\columnwidth]{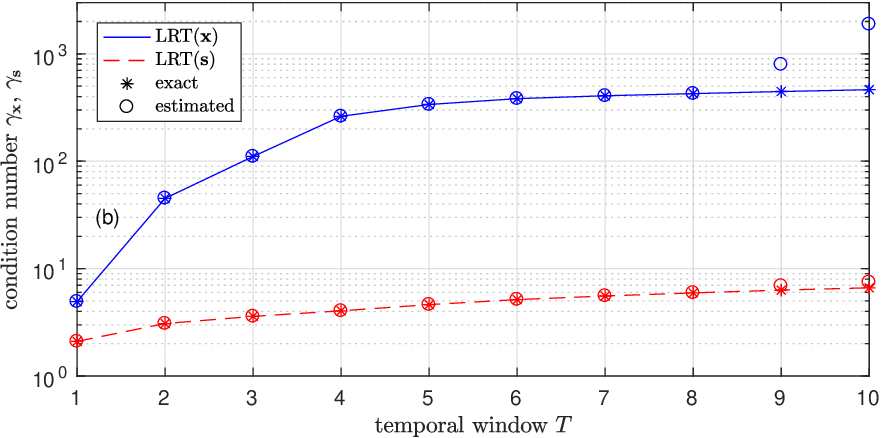}

  \vspace*{-.3cm}
  
  \caption{(a) Separability of distributions and (b) condition numbers of
    covariance matrices for setting with $J=20$ and
    the transient source 20~dB below the stationary sources.
    \label{fig:fig2}}
\end{figure}

Increasing the temporal correlation via $J=20$ and now dropping the
transient signal strength to match the noise floor,
Fig.~\ref{fig:fig2}(a) show the separability. Now over the range of
$T$, the LRT applied to the measurements is signifantly worse compared
to the LRT operating on the projected data.  Also note that for $T>8$,
the GLRT results on the measurements dramatically deteriorate. When
inspecting the involved matrices, it is not only that for e.g.~$T=10$,
the LRT involves the inversion of $100 \times 100$ matrices for the
measurement case as opposed to $30\times 30$ matrices in the case of
the projected data, but also that the condition numbers
$\gamma_{\Vektor{x},1}$ and $\gamma_{\Vektor{s},1}$ as defined in
Sec.~\ref{sec:cond} significantly deviate.  In case the condition
numbers are determined for the estimated covariance matrices, for
$T>8$ deviations become noticeable in Fig.~\ref{fig:fig2}(b), which
agrees with the performed drop for the measurement LRT in
Fig.~\ref{fig:fig2}(a).

%
%
\section{Discussion and Conclusions}

To detect a weak broadband transient signal, we have investigated the
application of a likelihood ratio test to polynomial
subspace-projected data rather than directly to the measurements.
While for the stronger transient signals, the latter approach is
optimal as it exploits all available information about the transient
signal, the restriction to the noise-only subspace in the case of
weaker transient signals had been shown to be beneficial. The subspace
project approach ideally suppresses any strong stationary source
components. This firstly removes any strong temporal correlations,
which hinder the test, and secondly also significantly reduce the
condition number of the covariance matrices that the LRT requires to
be inverted, hence providing. As a result, the subspace-projected LRT
operates on matrices that are decreased in both their dimension and
condition number. This effect is the more pronouced the stronger the
temporal correlations of the measured signals are and the weaker the
transient signal is compared to the stationary sources.

%
%

\end{document}

\bibliographystyle{IEEEtran}
\bibliography{/home/stephan/library/stephan_refAK_bib,
               /home/stephan/library/stephan_refLR_bib,
               /home/stephan/library/stephan_refSZ_bib}

\end{document}